\documentclass[reprint,superscriptaddress,amsmath,amssymb,aps,pra]{revtex4-2}

\usepackage{txfonts}
\usepackage{bm,dsfont,mathtools}
\usepackage{graphicx}
\usepackage{tikz}
\usepackage{physics}
\usepackage{xcolor}
\usepackage{amsmath}
\usepackage[colorlinks=true,breaklinks=true,allcolors=blue]{hyperref}

\newcommand{\im}{\mathrm{i}}

\makeatletter
\let\cat@comma@active\@empty
\makeatother

\begin{document}

\title{Quantum fluctuations and chaos in fully connected spin models}

\author{Aleksandra A. Ziolkowska}
\email{aaz1@st-andrews.ac.uk}
\affiliation{Institute of Physics,
		Johannes Gutenberg University Mainz, 
		  Staudingerweg 7, 
		55128 Mainz, Germany}
\affiliation{School of Physics and Astronomy,
        University of St Andrews,
        St Andrews KY16 9SS, 
        United Kingdom}
\author{Aleksandr N. Mikheev}
\email{aleksandr.mikheev@uni-konstanz.de}
\affiliation{Institute of Physics,
		Johannes Gutenberg University Mainz, 
		  Staudingerweg 7, 
		55128 Mainz, Germany}
\affiliation{Department of Physics,
		University of Konstanz, 
		  Universit{\"a}tsstra{\ss}e  10,
        78464 Konstanz, Germany}

\begin{abstract}
We investigate beyond-mean-field dynamics in a fully connected $\mathrm{SU}(3)$ spin-exchange model, focusing on the interplay between chaotic dynamics and quantum fluctuations. Using the two-particle irreducible (2PI) effective action formalism, we derive equations of motion that systematically account for higher-order correlations generated by interactions, and demonstrate how quantum fluctuations can regularize chaotic dynamics displayed by macroscopic observables. Our results show that an accurate treatment of fluctuations is essential for describing macroscopic dynamics in quantum many-body systems and promote 2PI as a robust framework for connecting microscopic correlations to macroscopic nonequilibrium phenomena.
\end{abstract}
\maketitle

\section{Introduction}
Fully connected models are one of the paradigmatic platforms for exploring nonequilibrium many-body physics \cite{Kitagawa:1993, Campbell2016, Kirton2019, Titum2020, Sehrawat2021}. These models naturally arise in quantum simulation platforms such as cavity-QED setups \cite{Ritsch:2012oze} or trapped-ion systems \cite{Monroe:2021,Foss-Feig:2024blk}, where programmable spin-spin interactions can interpolate between short-range and nearly infinite-range limits. In the thermodynamic limit, their dynamics is often captured by a small set of macroscopic variables, yet can still display nontrivial phenomena characteristic of strongly interacting systems. The success of mean-field theory in such models relies on the suppression of correlations with larger system size. However, modern quantum simulators increasingly operate in regimes where strong correlations can become important and probe observables that are explicitly sensitive to fluctuations, including two-time correlation functions, noise spectra, and response functions \cite{Ye_2013, Landig_2015, Garttner_2017, Wei_2022, Joshi_2025}. Understanding the effect of fluctuations is therefore essential for characterizing nonequilibrium phases realized in experiments and for assessing the robustness of phenomena such as superradiance \cite{Boneberg2022}, synchronization \cite{Xu_2014}, and time-crystalline order \cite{Heugel:2023}. Their role becomes especially important near dynamical instabilities, soft-mode regimes, and phase boundaries \cite{Nagy2011, Konya2012, Brennecke_2013}, where quantum fluctuations are not merely quantitative corrections but can qualitatively reshape phase diagrams and dynamical responses.

\begin{figure}[t]
	\centering
	\includegraphics[width=\linewidth]{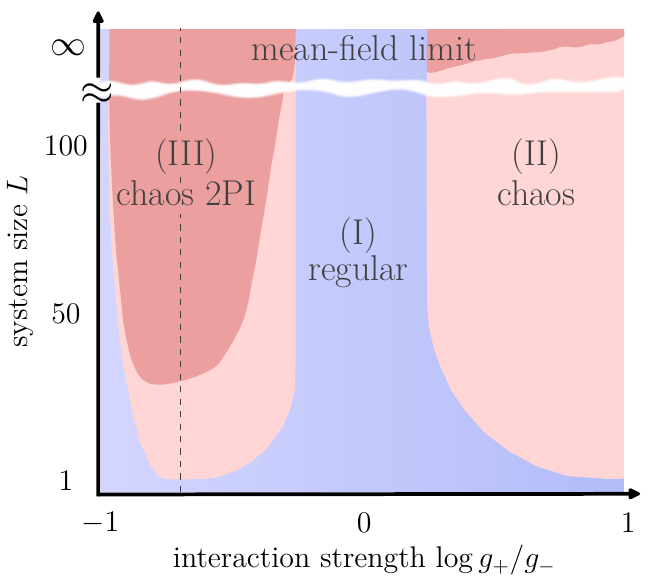}
	\caption{A sketch of the dynamical phase diagram for the $\mathrm{SU}(3)$ spin-exchange Hamiltonian initialized in a chaotic state for fixed $h = g_{-} = 1$. All parameters refer to Eq.~\eqref{eq:hamiltonian} in the main text. Phase (I) represents regular dynamics, phase (II) chaotic dynamics as indicated by the second-order cumulant expansion, and phase (III) designates chaotic dynamics as shown by the 2PI computations with next-to-leading order corrections. The upper limit of the graph with an infinite system size corresponds to the mean-field limit, which is the same in both approaches. The dashed line corresponds to the ratio $g_{+}/g_{-} = 0.2$ analyzed in Fig.~\ref{fig:dT_chaos}. See Sec.~\ref{sec:chaos} for the detailed discussion of the phase diagram.}
	\label{fig:phase_diagram}
\end{figure}

A common approach to incorporating beyond-mean-field effects is the cumulant expansion, in which the hierarchy of operator moments is truncated by neglecting higher-order correlations \cite{Kubo:1962, VanKampen:1974, gardiner2004handbook}. It provides a flexible and transparent approximation scheme that, even in its simplest form, can often show good agreement with exact dynamics across a broad range of settings \cite{Kirton:2017bly,Plankensteiner:2021wst}. At the same time, it is known that fixed-order cumulant truncations are generally uncontrolled and can exhibit secular behavior or poor convergence when strong correlations quickly build up \cite{ Rajagopal:1974, Schack:1990, Pavskauskas:2012, Sanchez-Barquilla:2020pug,Fowler_Wright_2023, Kerber:2025}. While several extensions of the cumulant framework have been designed to alleviate these issues  \cite{ Reichman1997, Kira2008, Bertini:2015, Robinson:2022_2, Szankowski:2022eol}, they usually introduce additional technical complexity and often require problem-specific choices, motivating complementary frameworks that organize correlation effects in a systematic and self-consistent manner. In this work, we adopt the two-particle irreducible (2PI) effective action formalism, highlighting its potential as a systematic and self-consistent approach to beyond-mean-field dynamics. This functional path-integral approach is based on selective resummations of interaction effects, yielding self-consistent equations for nonequilibrium Green’s functions that incorporate the feedback of higher-order correlations \cite{Danielewicz:1982kk,Berges:2004yj}. It is therefore especially well suited to far-from-equilibrium problems in which higher-order correlations quickly build up dynamically. 

To showcase the utility of the 2PI framework, we investigate how fluctuations modify, suppress, and reshape chaotic macroscopic dynamics in a fully connected $\mathrm{SU}(3)$ spin-exchange model. As illustrated in Fig.~\ref{fig:phase_diagram}, the chaotic dynamics of the macroscopic spins is regularized by strong fluctuations, which arise either at strong interaction strengths, corresponding to the right side of the diagram, or at small system sizes, corresponding to the lower part of the diagram. In addition, the phase diagram contains a regular region near its center, associated with the vicinity of an integrable point where the dynamics effectively reduces to $\mathrm{SU}(2)$. While the 2PI framework captures these fluctuation-induced regularization effects through self-consistently dressed two-point functions, the corresponding low-order cumulant truncation misses correlation feedback in the strong-interaction regime. A detailed discussion of the phase diagram and the underlying mechanisms is given in the main text.

The paper is organized as follows. We begin in Sec.~\ref{sec:model} by introducing the model under consideration. In Sec.~\ref{sec:formalism}, we discuss the evolution equations governing the dynamics of the system’s observables. Starting from the conventional low-order cumulant expansion, we discuss the role of higher-order correlations that emerge dynamically during the evolution and the importance of treating them self-consistently. This motivates the use of the two-particle-irreducible (2PI) formalism, which we briefly review before presenting the $1/L$ expansion for fully connected systems. In Sec.~\ref{sec:chaos}, we use these methods to study how fluctuations affect the chaotic dynamics of macroscopic observables. We summarize and draw our conclusions in Sec.~\ref{sec:conclusions}. 

\section{Model}
\label{sec:model}
We consider a fully connected spin-exchange model of $\mathrm{SU}(N)$ spins. As discussed in the introduction, such models naturally emerge as effective descriptions of $N$-level systems uniformly coupled to a common bosonic mediator, such as a cavity photon mode in optical or circuit QED \cite{Morrison:2008exn,Larson:2010kxo,ValenciaTortora:2023} or collective phonon modes in trapped-ion systems \cite{Porras:2004ekq,Korenblit:2012fqk}. In the far-detuned regime, the mediator can be adiabatically eliminated, resulting in virtual exchange interactions between all pairs of spins. The simplest case $N=2$ recovers conventional collective $\mathrm{SU}(2)$ spin models, with the Lipkin--Meshkov--Glick Hamiltonian \cite{Lipkin:1964yk,Meshkov:1965btx,Glick:1965} serving as a paradigmatic example. This class of models is known to exhibit only regular dynamical responses, such as relaxation and persistent oscillations \cite{ValenciaTortora:2023}.

The absence of complex, irregular dynamical regimes in the $\mathrm{SU}(2)$ case can be traced to the limited number of degrees of freedom, $n=3$, in the effective collective-spin description. Combined with energy and spin conservation, this strongly restricts the possible range of dynamical responses. The minimal fully connected spin-exchange model capable of chaotic dynamics is consequently realized by $\mathrm{SU}(3)$ spins. In this work, we consider a bosonic realization of such a fully connected spin model described by the Hamiltonian
\begin{align}
\label{eq:hamiltonian}
\hat{H}
&=
h\sum_{j=1}^L\left(\hat{T}_{j,11} - \hat{T}_{j,33}\right)\nonumber\\
&-
\frac{1}{L}\sum_{i,j=1}^L\left(g_+\hat{T}_{i,12} + g_-\hat{T}_{i,23}\right)\left(g_+\hat{T}_{j,21} + g_-\hat{T}_{j,32}\right),
\end{align}
where $h$ is a homogeneous magnetic field and $g_\pm$ are the hopping rates corresponding to each of the allowed transitions between levels $\alpha\in\{1,2,3\}$. The operators $\hat{T}_{j,\alpha\beta} = b^\dagger_{j,\alpha}b_{j,\beta}$, defined as bosonic bilinears, form effective $\mathrm{SU}(3)$ spin operators, satisfying the algebra
\begin{equation}
\label{eq:algebra}
\big[\hat{T}_{i,\alpha\beta},\hat{T}_{j,\gamma\delta}\big] = \delta_{ij}\left(\delta_{\beta\gamma}\hat{T}_{j,\alpha\delta}-\delta_{\delta\alpha}\hat{T}_{j,\gamma\beta}\right) \equiv \mathrm{i} \delta_{ij} f_{\alpha\beta;\gamma\delta;\epsilon\eta} \hat{T}_{j,\epsilon\eta}.
\end{equation}
Here and in the following, the standard Einstein summation convention over repeated indices is implied. 

The above model may be viewed as a 3-level generalization of the Tavis--Cummings model with an adiabatically eliminated photonic mode. Previous works have shown that a departure from the $\mathrm{SU}(2)$ regime to a higher-dimensional local Hilbert space gives rise to new dynamical responses. In particular, the mean-field analysis indicates that the $\mathrm{SU}(3)$ model can exhibit chaotic dynamics \cite{ValenciaTortora:2023, Balducci:2025}, while the spectral diagnostics in the quantum limit revealed symmetry sectors with GOE level spacing, indicative of quantum chaos \cite{Balducci:2025}. In this respect, the present work aims to bridge these two limits by examining how quantum fluctuations affect the mean-field dynamical responses.

\section{Equations of Motion}
\label{sec:formalism}

\subsection{Second-order cumulant expansion}
\label{sec:2pt_cumulants}
A natural starting point for describing quantum dynamics is to consider the Heisenberg equations of motion for expectation values of observables:
\begin{equation}
\mathrm{i}\frac{\dd}{\dd t}\,\big\langle\hat{\mathcal{O}}(t)\big\rangle = \big\langle\big[\hat{\mathcal{O}},\hat{H}\big]\big\rangle\,.
\end{equation}
Here, $\hat{\mathcal{O}}$ could be, for instance, either a bosonic operator $\hat{b}_{j,\alpha}$ or a spin operator $\hat{T}_{i,\alpha\beta}$. In general, the right-hand side of this equation will involve expectation values of higher-order products of the operator $\hat{\mathcal{O}}$, thus forcing one to include equations for higher-order correlation functions $\langle \hat{\mathcal{O}}_1 \hat{\mathcal{O}}_2 \rangle$, $\langle \hat{\mathcal{O}}_1 \hat{\mathcal{O}}_2 \hat{\mathcal{O}}_3  \rangle$, etc. While the complexity of the problem can be reduced through the use of conservation laws and symmetries, this usually gives rise to an infinite BBGKY-type hierarchy of equations coupling the dynamics of correlation functions of all orders. 

A natural approach to circumvent this problem consists of truncating the hierarchy at a given finite order. This can be achieved by systematically decomposing higher-order correlation functions into products of lower-order correlations, a procedure known as the cumulant expansion. At lowest order, this procedure yields the standard mean-field approximation, while the next order involves two-point correlations, thereby incorporating Gaussian fluctuations. In fact, even in nonequilibrium settings, much of the relevant information about the system is often encoded in two-point correlation functions. For example, the diagonal entries of the correlation function $\langle\hat{b}_{\alpha}^{\dagger} \hat{b}_{\beta}\rangle$ provide information about the average number of bosons occupying a given atomic level, while the off-diagonal components describe coherences between the respective levels. It is therefore natural to seek an effective description for the dynamics of two-point correlation functions.

For the model under consideration, the second-order cumulant expansion admits a transparent physical interpretation in terms of semiclassical equations of motion for the $\mathrm{SU}(3)$ effective spin degrees of freedom, which can be derived in the basis of the $\mathrm{SU}(3)$ coherent states \cite{Davis2019Photon,Dahlbom2022Langevin}. Defining collective $\mathrm{SU}(3)$ spin operators $\hat{T}_{\alpha\beta} \equiv \sum_{j=1}^L\hat{T}_{j,\alpha\beta}$, and their average expectation values $T_{\alpha\beta}=\expval{\hat{T}_{\alpha\beta}}/L = \sum_{j=1}^L \expval{\hat{b}^{\dagger}_{j,\alpha} \hat{b}_{j,\beta}}/L$, evaluated in the subspace spanned by $\mathrm{SU}(3)$ coherent state, the corresponding classical equations of motion assume the Poisson--Lie form
\begin{equation}
\label{eq: classicalEom}
\frac{\dd}{\dd t}T_{\alpha\beta} = f_{\alpha\beta;\gamma\delta;\epsilon\eta}\ \frac{\partial \mathcal{H}}{\partial T_{\gamma\delta}}T_{\epsilon\eta}\,,
\end{equation}
where $f_{\alpha\beta;\gamma\delta;\epsilon\eta}$ are the structure constants introduced in Eq.~\eqref{eq:algebra} and $\mathcal{H}$ is the classical Hamiltonian functional, as detailed in App.~\ref{app:cumulants}.

Notably, the Hamiltonian \eqref{eq:hamiltonian} can be written entirely in terms of the collective spin operators $\hat{T}_{\alpha\beta}$, which obey a modified algebra $[\hat{T}_{\alpha\beta},\hat{T}_{\gamma\delta}] = \mathrm{i} L^{-1} \delta_{ij} f_{\alpha\beta;\gamma\delta,\epsilon\eta} \hat{T}_{\epsilon\eta}$. This suggests that the inverse system size plays the role of an effective Planck constant, $\hbar_{\mathrm{eff}} = 1/L$, controlling the strength of fluctuations in the system, as is typical for all-to-all connected models. Using $1/L$ as the control parameter, it is possible to include the next order corrections to the classical equations of motion. We refer to App.~\ref{app:cumulants} for more details on the derivation.

\subsection{Effect of higher-order correlations}
\label{sec:higher_order}
The second-order cumulant equations of motion introduced in the previous section ignore the effect that the build-up of higher-order correlation functions has on the two-point functions. This approximation is expected to quickly break down in the presence of strong fluctuations, where higher-order cumulants become large and cannot be neglected.

Within the cumulant expansion approach, these effects are included by truncating the hierarchy at higher order, thereby incorporating higher-order (connected) correlation functions such as $\langle\hat{T}_{\alpha\beta} \hat{T}_{\gamma\delta}\rangle_c$, $\langle\hat{T}_{\alpha\beta} \hat{T}_{\gamma\delta} \hat{T}_{\epsilon\eta}\rangle_c$, and so forth, resulting in a local-in-time description of the dynamics. Instead, one should expect the effective description of the dynamics to acquire non-local effects when eliminating the higher-order degrees of freedom. As a simple example illustrating this point, consider a system of two coupled differential equations
\begin{equation}
\label{eq:toy_integrated}
\dot{x}(t) = a(t)\,y(t)\,,\quad \dot{y}(t) = b(t)\,x(t)\,,
\end{equation}
where $x(t)$ and $y(t)$ represent two-point and higher-order correlation functions, respectively. Eliminating $y(t)$ leads to an integro-differential equation for the remaining degree of freedom
\begin{equation}
\dot{x}(t) = a(t) \left[y(t_0)+ \int_{t_0}^t \dd{t'} b(t')\,x(t') \right].
\end{equation}
Accordingly, one concludes that any method that does not include the infinite tower of differential equations dictated by the BBGKY hierarchy implicitly ``integrates out'' the higher-order correlation functions, with their effects entering through an effective memory kernel ($b(t)$ in the above toy example).
 
Importantly, as will be further elaborated below, for the resulting dynamics to be internally consistent, the effective memory kernel must have a specific structure \cite{Baym:1961}. This explains why approaches based on the conventional cumulant expansion approximation are often non-conserving and prone to secular behavior, since a generic truncation of the hierarchy does not necessarily yield the correct structure of the effective memory kernel. In the following section, we outline a field-theoretic approach that provides a systematic prescription for a memory kernel that ensures a self-consistent dynamics in terms of two-point correlation functions.  

\subsection{2PI formalism}
We begin by formally writing the Dyson equation for the time-ordered, two-point connected correlation functions $G_{ij}^{\alpha\beta}(t,t')=\expval{\mathcal{T} \hat{b}_{i,\alpha}(t)\,\hat{b}_{j,\beta}^\dagger(t')}_c$, which reads
\begin{equation}
\label{eq:Dyson}
G = G_0 + G_0 \circ \Sigma \circ G\,.
\end{equation}
Here, $\circ$ denotes a generalized convolution that sums and integrates over all possible indices and coordinates. This exact identity relates the full (interacting) propagator $G$ to the bare (non-interacting) propagator $G_0$ and the self-energy $\Sigma$, which encodes the renormalization of the propagator due to interactions, capturing phenomena such as frequency renormalization, damping, and collective excitations. 

Multiplying both sides by $G_0^{-1}$ one can rewrite the Dyson equation \eqref{eq:Dyson} in a slightly different form:
\begin{equation}
\label{eq:Dyson2}
G_0^{-1} \circ G = \mathds{1} + \Sigma \circ G\,. 
\end{equation}
As the bare inverse propagator $G_0^{-1}$ normally takes the form of a linear differential operator, the above equation is structurally very similar to Eq.~\eqref{eq:toy_integrated}, with the self-energy $\Sigma$ playing the role of the memory kernel. It encodes the information about the structure of correlations induced by the nonlinearities and diagrammatically represents an infinite sum of one-particle irreducible diagrams with two amputated legs \cite{Weinberg:1995}. Here, $n$-particle irreducible ($n$PI) means that the diagram cannot be decomposed into two disconnected parts by cutting at most $n$ propagator lines. 

\begin{figure}[t]
\centering
\includegraphics[width=0.9\linewidth]{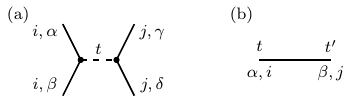}
\caption{Diagrammatic notation. (a) A 4-point vertex describing the interaction of two bosons on site $i$ with two bosons on site $j$ occurring at time $t$. The interaction strength  $g_a g_b/L$, with $a, b \in \{+,-\}$, depends on the type of bosons entering the interaction, cf. Eq.~\eqref{eq:hamiltonian}. (b) Two-point correlation function $G_{ij}^{\alpha\beta}(t,t')$. If the on-site $\mathrm{U}(1)$ symmetry of the Hamiltonian \eqref{eq:hamiltonian} is not broken, the two-point function is diagonal in coordinate space, $G_{ij}^{\alpha\beta}(t,t') \propto \delta_{ij}$.}
\label{fig:notation}
\end{figure}

The self-energy is generally intractable without a truncation or approximation. The standard perturbative approach consists of expanding the self-energy $\Sigma$ to finite order using the non-interacting propagator $G_0$. Out of equilibrium, however, this procedure typically gives rise to secular terms that grow as powers of time, ultimately leading to a breakdown of the dynamics \cite{Berges:2004yj}. As first argued in Ref.~\cite{Baym:1961}, the key to resolving this problem is to employ, in the expansion of the self-energy $\Sigma$, the same interacting propagator $G$ that appears on the left-hand side of Eq.~\eqref{eq:Dyson2}. Mathematically, this is achieved if the self-energy has the form $\Sigma \sim \delta \Gamma_2[G]/\delta G$, where $\Gamma_2$ is the sum of two-particle irreducible, with respect to the full propagator $G$, connected vacuum diagrams. One can conveniently frame this prescription as a variational principle $\delta \Gamma_{\mathrm{2PI}}/\delta G = 0$ upon introducing the functional
\begin{equation}
\Gamma_{\mathrm{2PI}} = S + \frac{\im}{2} \Tr\ln G^{-1} + \frac{\im}{2} \Tr G_0^{-1}\,G + \Gamma_2\,,
\end{equation}
which goes by the name of the two-particle irreducible (2PI) effective action, also known as the Luttinger--Ward \cite{Luttinger:1960} or the Cornwall--Jackiw--Tomboulis \cite{Cornwall:1974} potential. Here, $S$ is the classical action, and the proper self-energy is defined as $\Sigma[G] = 2\mathrm{i}\delta \Gamma_2[G]/\delta G$ in this convention. 

\subsection{$1/L$ expansion of the 2PI effective action}
Naturally, when all diagrams are included, both the standard perturbative expansion in terms of the bare propagators $G_0$ and the 2PI approach described in the previous sections are equivalent and yield the exact solution. A key feature of the 2PI formalism, however, is that it guarantees internal self-consistency of the resulting dynamical equations even when the expansion of $\Gamma_2$ is truncated at a finite order. As always, the choice of diagrams used to approximate this series is not unique and usually consists of a systematic expansion in some small parameter, the most popular choice of which is the interaction coupling constant. However, as mentioned at the end of Sec.~\ref{sec:2pt_cumulants}, fully connected models offer an alternative expansion parameter that controls the strength of fluctuations, namely the inverse system size $1/L$, allowing one to go beyond the weak-coupling limit in a systematic manner.

\begin{figure}[t]
\centering
\includegraphics[width=\linewidth]{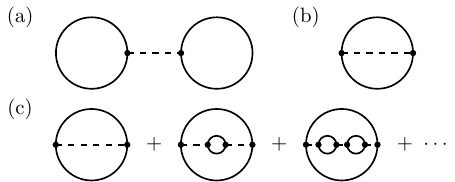}
\caption{Diagrams contributing to the 2PI effective action considered in this work. (a) The only diagram contributing at leading order $\mathcal{O}(L)$ in the $1/L$ expansion. (b) Time-local diagram contributing at next-to-leading order $\mathcal{O}(L^{0})$. (c) An infinite series of next-to-leading-order diagrams. These include the time-local diagram (b) as well as an infinite number of time-nonlocal processes.}
\label{fig:diagrams}
\end{figure}

To see how the large-$L$ expansion works within the 2PI formalism, we first introduce in Fig.~\ref{fig:notation} the graphical notation for the building blocks that constitute the diagrammatic expansion of the 2PI effective action. Starting with the first order in the interaction, there are two distinct ways to close the propagator lines, as depicted in Figs.~\ref{fig:diagrams}(a) and (b), with their contributions schematically reading as $G_{ii} G_{jj}$ and $G_{ij} G_{ij}$, respectively. Provided the on-site $\mathrm{U}(1)$ symmetry of the Hamiltonian is not broken, the propagators are diagonal in coordinate space, $G_{ij} \propto \delta_{ij}$. Combined with the $1/L$ prefactor from the interaction vertex, cf. Eq.~\eqref{eq:hamiltonian}, it can be deduced that the diagram in Fig.~\ref{fig:diagrams}(a) scales as $\mathcal{O}(L)$, whereas the diagram Fig.~\ref{fig:diagrams}(b) is suppressed by a factor of $1/L$ compared to it, scaling only as $\mathcal{O}(L^0)$. One refers to such contributions as leading-order (LO) and next-to-leading-order (NLO) in $1/L$, respectively. 

Both diagrams are local in time, implying that the corresponding contributions to the self-energy are likewise time-local, $\Sigma^{(a/b)}(t,t') \propto \delta(t-t')$. As a result, the $\Sigma \circ G$ term in Eq.~\eqref{eq:Dyson2} remains local-in-time at first order in the interaction. At the same time, an analysis of diagrams with multiple vertices reveals an infinite series of time-nonlocal ``bubble-chain'' diagrams that all contribute at next-to-leading order in the $1/L$ expansion, see Fig.~\ref{fig:diagrams}(c). In fact, one can show that this is the only class of diagrams that contributes at $\mathcal{O}(L^0)$. Remarkably, the bubble-chain diagrams can be efficiently resummed into a geometric series or mapped to a two-loop diagram by introducing an auxiliary field via the Hubbard--Stratonovich transform, yielding a closed analytic expression for $\Gamma_2$ and, consequently, for the self-energy $\Sigma$. 

Although nonequilibrium Green's functions provide insight also to unequal-time correlations, in this work we focus on the time-diagonal entries of the two-point functions. Appropriately adjusted, the latter reduce to the expectation values of the effective $\mathrm{SU}(3)$ spin operators. Using the NLO nonequilibrium Dyson equations, their evolution is described by the equation of the form
\begin{equation}
\label{eq:KB}
\mathrm{i}\frac{\dd}{\dd t}T(t)=[T(t),\Tilde{M}(t)]  + \mathrm{memory~terms}\,.
\end{equation}
Here, the first term can be identified as a local-in-time contribution arising from the second-order cumulant expansion, while the  memory terms involve a convolution of unequal-time Green’s functions with a time-nonlocal self-energy $\Sigma$. It is worth emphasizing again that this term plays a central role in the relaxation dynamics of the system. For instance, within the kinetic theory approximation, it is the memory terms that lead to the collision integral of the Boltzmann equation and therefore account for key relaxation processes such as energy redistribution within the system. The second-order cumulant expansion entirely misses this crucial physics. At the same time, higher-order cumulant truncations, although capable of incorporating such relaxation processes, generally result in dynamics that is not internally self-consistent, potentially leading to unstable and secular behavior, as discussed at the end of Sec.~\ref{sec:higher_order}. 

In the remainder of this paper, we analyze how these  memory terms modify the dynamical responses of the fully connected $\mathrm{SU}(3)$ spin model defined in Eq~\eqref{eq:hamiltonian}, with particular emphasis on the role of fluctuations in regularizing chaotic dynamics displayed by the macroscopic observables. We refer to App.~\ref{app:2PI} for details on the derivation of the NLO 2PI effective action, the corresponding self-energy, and the respective nonequilibrium Dyson equations.

\section{Chaos in the presence of quantum fluctuations}
\label{sec:chaos}
\subsection{Preliminary considerations}
Let us start by examining the ratio $g_{+}/g_{-}$ of the two interaction strength parameters for the $\mathrm{SU}(3)$ model. For $g_{+}/g_{-}=1$, the system becomes integrable, while in the limits $g_{+}/g_{-} \to 0$ and $g_{+}/g_{-} \to \infty$ the dynamics becomes confined to a smaller part of the parameters' space. Away from these limiting cases, one expects a competition between chaos and fluctuations, whose magnitude is controlled by the system size $L$ and the interaction strength with respect to the magnetic field $h$. We use the latter as the energy scale in which all other parameters are measured, setting $h=1$ accordingly. To further reduce the number of control parameters, we have fixed $g_{-} = 1$, so that the overall interaction strength is entirely determined by the remaining coupling $g_{+}$.  We emphasize that the above parameters constitute only a part of the full parametric space of the model and were chosen to underline the key interplay between chaos and quantum fluctuations. 

As a representative initial condition, we consider a state that features an unbalanced population distribution among the bosonic levels, thereby avoiding spurious symmetries and allowing for chaotic dynamics. More concretely, we have chosen as the initial condition a direct product of identical coherent states at each site,
\begin{equation}
    \ket{\psi_0} = \bigotimes_{j=1}^L \frac{1}{\sqrt{A_1^p A_2^q}}\,\mathrm{e}^{\gamma_3 \hat{T}_{j,31}}\mathrm{e}^{\gamma_1 \hat{T}_{j,21}} \mathrm{e}^{\gamma_2 \hat{T}_{j,32}}\ket{\mu}_{p,q}\,,
\end{equation}
with $p=2$, $q=1$, corresponding to $\mathcal{N}_a=4$ particles per site, and free parameters $\gamma_1=4/\sqrt{21}$, $\gamma_2 = 2/\sqrt{21}$, and $\gamma_3 = \mathrm{i}/\sqrt{21}$. Here, $\ket{\mu}_{p,q}$ is the highest weight state of the $D(p,q)$ representation of $\mathrm{SU}(3)$ and the normalization constants are given by $A_1 = 1 + |\gamma_1|^2 + |\gamma_3|^2$, $A_2 = 1 + |\gamma_2|^2 + |\gamma_3 -\gamma_1\gamma_2|^2$, see App.~\ref{app:coherent} for more details. 

In general, chaotic systems exhibit mixed phase spaces that can be partitioned into distinct basins of regular and chaotic motion, whose relative size and structure change as the Hamiltonian's parameters change \cite{Hutt2020}. A comprehensive study of the Hilbert space structure of this model and its significance for the classical phase-space structure can be found in Ref.~\cite{Balducci:2025}.

\subsection{Dynamical phase diagram}
We diagnose chaos through the time-evolution of the expectation values of the collective $\mathrm{SU}(3)$ spins, investigating how the dynamical responses of the ``trajectories'' $T_{\alpha\beta}(t)$ differ as a function of the interaction strength and the system size $L$. Chaotic dynamics is characterized by an exponential divergence of different trajectories at early times, $\Delta T_{\mathrm{chaotic}}(t)\sim \mathrm{e}^{\lambda t}$, where $\lambda$ is the maximum Lyapunov exponent. As a measure of the distance between trajectories, we used the Frobenius norm defined as 
\begin{equation}
\label{eq:dT}
\Delta T(t)=\frac{2}{R(R-1)}\sum_{i>j}\sqrt{\sum_{\alpha,\beta=1}^3 \left|T_{\alpha\beta}(t;i)-T_{\alpha\beta}(t;j)\right|^2}\ ,
\end{equation}
where $i,j=1,\dots,R$ enumerate the differently perturbed initial conditions and $R$ indicates the number of initial conditions sampled. 

We emphasize that the notion of trajectories used in this work should not be confused with the microscopic trajectories employed in methods such as the truncated Wigner approximation (TWA). In the latter, microscopic trajectories correspond to solutions of the classical equations of motion whose initial conditions are sampled from the quasiprobability distribution (wave packet) associated with the initial quantum state. The ensuing dynamics of these trajectories describes the propagation and diffusion of the wave packet in phase space. By contrast, the macroscopic trajectories introduced here represent the time evolution of the wave packet’s mean position. When fluctuations are weak, this evolution is largely governed by the underlying classical dynamics and therefore reflects its regular and chaotic regimes. Conversely, strong fluctuations lead to diffusion of the wave packet, reflected by the growth of higher-order correlations such as  $\langle\hat{T}_{\alpha\beta} \hat{T}_{\gamma\delta}\rangle_c$ and $\langle\hat{T}_{\alpha\beta} \hat{T}_{\gamma\delta} \hat{T}_{\epsilon\eta}\rangle_c$, effectively washing out the signatures of underlying classical chaos. This again underscores the role of the memory terms in shaping the dynamical responses of the macroscopic observables, since it is precisely the memory part $\Sigma \circ G$ that encodes information about higher-order correlations, as elaborated in Sec.~\ref{sec:formalism}. For further discussion on the smoothing role of fluctuations on classical chaos, we refer to the literature \cite{Casetti:1998,Matinyan:1997,Altland:2012,Altland_2012_Dicke}.

In Fig.~\ref{fig:phase_diagram}, we present a sketch of the dynamical phase diagram for the $\mathrm{SU}(3)$ spin exchange model, as diagnosed through the temporal dependence of $\Delta T(t)$. A chaotic response is signaled by exponential growth of the norm, while   regular dynamics are characterized by a slower, polynomial growth. Close to the integrable points $g_+\rightarrow 0$ and $g_+=g_-$, where the Hamiltonian reduces to a representation of $\mathrm{SU}(2)$ atoms, the system exhibits regular dynamics, which we mark as dynamical phase I. Moving away from the integrable limit, in phase II, collective $\mathrm{SU}(3)$ spins display chaotic dynamics at the level of the second-order cumulant expansion, while strong fluctuations captured by the NLO 2PI approximation regularize this behavior. The discrepancy between the methods is structural: in a chaotic system, correlations generated at low order are rapidly transferred to higher moments, so any fixed-order closure generically underestimates the backaction of fluctuations on macroscopic dynamics. By contrast, the 2PI formalism includes those effects self-consistently, while retaining a description in terms of two-point functions. Notably, although second-order cumulants capture the effects of system size to a limited extent, they are evidently largely insensitive to the interaction strength. As a result, the cumulant equations of motion show a chaotic response even in the bottom-right corner of Fig.~\ref{fig:phase_diagram}, where strong fluctuations arising from both the small system size and the strong interaction strength are expected to regularize the dynamics. Finally, in dynamical phase III, the underlying classical chaos is sufficiently strong to persist even in the presence of fluctuations induced by the infinite chain of rescattering processes captured by the NLO 2PI approximation. In this regime, interaction-induced fluctuations merely soften the exponential growth of diverging trajectories, reducing the maximal Lyapunov exponent in $\Delta T(t) \sim \mathrm{e}^{\lambda t}$ without altering its functional form. Note that phases II and III overlap, as the cumulant approach also reproduces chaotic behavior in that parameter range. In addition, both methods converge to the same mean-field description in the thermodynamics limit, showcasing a mostly chaotic response in the parameter regime away from integrability.

\begin{figure}[t]
    \centering
    \vspace{0.5ex}
    \includegraphics[width=\linewidth]{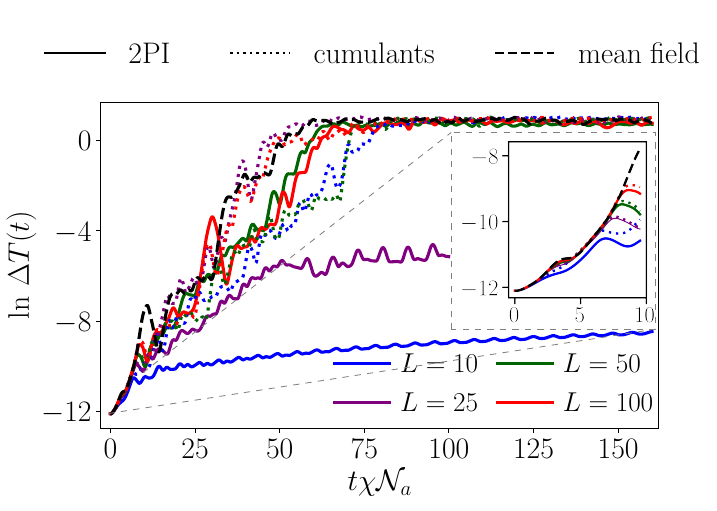}
    \caption{Time dependence of the logarithm of the Frobenius-norm distance between trajectories for chaotic initial conditions and Hamiltonian parameters $h=1$, $g_{+}=0.2$, and $g_{-}=1$. We compare three system sizes within the NLO 2PI, the second-order cumulant expansion, and the mean-field approximations. }
    \label{fig:dT_chaos}
\end{figure}

We further investigate the influence of system size on the dynamics by taking a cut across the phase diagram at $g_+/g_-=0.2$ and plotting the time dependence of the trajectory divergence $\Delta T$. In Fig.~\ref{fig:dT_chaos}, we show the logarithm of the divergence between trajectories as a function of the rescaled time $t\chi\mathcal{N}_a$, with $\chi = g_{+} g_{-}$. All chaotic trajectories are expected to saturate to the same value, indicative of trajectories reaching the typical distance allowed by the phase space. Any growth of the Frobenius norm that is slower than exponential indicates regular dynamics, for which the distance between trajectories also saturates at a lower value, consistent with the lack of ergodicity in nonchaotic dynamics. It is important to note that the the time dependence of the trajectories' distance in Fig.~\ref{fig:dT_chaos} is strongly shaped by the onset of quantum fluctuations, which is a timescale defined by the first departure from the mean-field-like behavior. As indicated by the curves computed using the NLO 2PI framework, this departure happens earlier for small system sizes, when the system is in the strongly correlated regime. Although both the 2PI and the cumulant expansion equations of motion converge towards the mean-field behavior for large system sizes, the cumulant expansion consistently underestimates the magnitude of fluctuations and hence adheres to the mean-field curve for longer times, as highlighted in the inset of Fig.~\ref{fig:dT_chaos}.  

\begin{figure}[t]
\centering
\includegraphics[width=\linewidth]{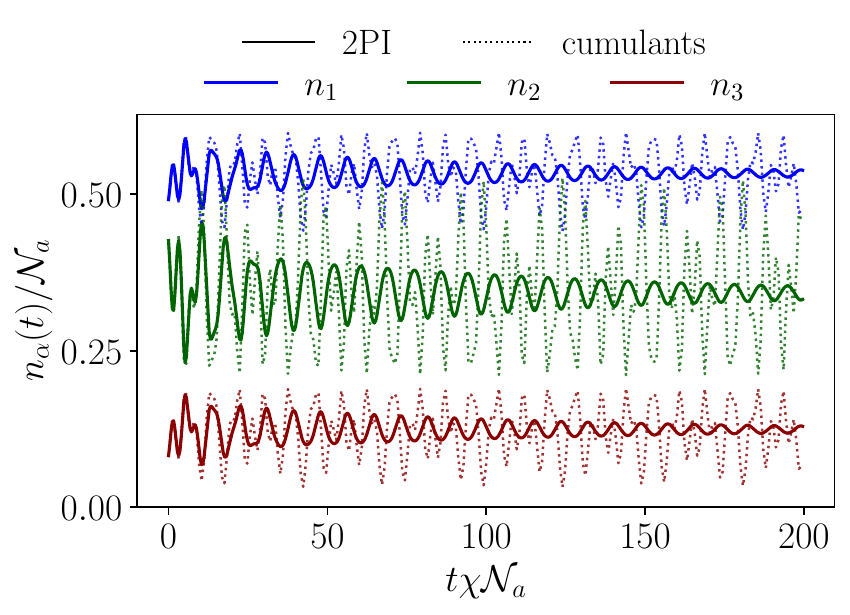}
\caption{Time dependence of the normalized bosonic level populations for $h=1, g_+=2, g_-=1$ and the system size $L=15$. The NLO 2PI approximation captures relaxation of the populations, whereas the second-order cumulant expansion exhibits persistent oscillations.}
\label{fig:observables}
\end{figure}

To highlight the decisive role of fluctuations in the dynamics of macroscopic observables, we finally show in Fig.~\ref{fig:observables} the evolution of bosonic level populations for individual trajectories. Although the shown dynamics corresponds to a chaotic initial condition, the second-order cumulant expansion does not capture the expected relaxation of the bosonic populations. This failure can be traced back to the weak-non-Gaussianity requirement inherent in the standard cumulant expansion framework. On the other hand, the equilibration of chaotic quantum systems can be understood as the progressive breakdown of Gaussianity.  Low-order cumulant expansions fail in this regime because they rely on assuming that higher-order cumulants remain small, whereas, in the chaotic system, they grow exponentially and dominate the relaxation process. For completeness, in App.~\ref{app:coherences}, we show the time dependence of the coherences corresponding to the populations in Fig.~\ref{fig:observables}.

\section{Conclusions}
\label{sec:conclusions}
In this work, we investigated the chaotic-to-regular dynamical transition in a collective $\mathrm{SU}(3)$ spin model. As the strength of fluctuations in such models is controlled by both the interaction strength and the system size, they provide a natural way to tune fluctuation effects. Our results show that fluctuations can substantially reshape the dynamics of macroscopic observables in quantum many-body systems. In regimes that are classically chaotic, these fluctuations may regularize the evolution by smoothing out otherwise chaotic trajectories over a broad region of parameter space. This regularizing effect is especially pronounced in the strong-fluctuation regime, where the build-up of correlations rapidly washes away classical, mean-field-type behavior even for initially (semi)classical states.

Within conventional local-in-time approaches, such as cumulant-expansion approximations, capturing these effects generally requires high-order truncations, since low-order schemes can miss the relevant physics. In contrast, the 2PI effective action formalism offers a robust framework for incorporating fluctuations beyond mean-field theory in a self-consistent way using only two-point correlation functions. Through memory kernels in the evolution equations, it naturally accounts for the feedback of dynamically built-up correlations without explicitly introducing higher-order correlators into the description. Despite its long-standing use in nonequilibrium condensed matter theory and high-energy physics, the 2PI framework remains comparatively underused in atomic and optical quantum systems. Our results promote it as a viable complementary theoretical framework for studying dynamics in modern quantum-simulation platforms operating in strongly correlated regimes, where rapid growth of correlations makes conventional mean-field-type approaches harder to control and their systematic extension to higher orders increasingly challenging.

\section*{Acknowledgements}
The authors thank H. Hosseinabadi and J. Marino for fruitful discussions and for their involvement in the earlier stages of the project. We further thank J. Marino for providing the inspiration to investigate the topic considered in this work. A.A.Z. acknowledges the support of the Alexander von Humboldt Foundation for the duration of this project. A.N.M. acknowledges financial support by the Swiss National Science Foundation (SNSF) through Sinergia Grant No. CRSII5 206008/1. This project was supported by QuantERA II Programme that has received funding from the European Union’s Horizon 2020 Research and Innovation Programme under Grant Agreement No 101017733 (``QuSiED'') and by the Dynamics and Topology Center funded by the State of Rhineland Palatinate. We gratefully acknowledge the computing time granted through the project ``DysQCorr'' on the Mogon II supercomputer of the Johannes Gutenberg University Mainz (\href{https://hpc.uni-mainz.de/}{hpc.uni-mainz.de}), which is a member of the AHRP (Alliance for High Performance Computing in Rhineland Palatinate, \href{http://www.ahrp.info/}{www.ahrp.info}), and the Gauss Alliance e.V. The authors acknowledge the use of the AI-based language model ChatGPT (by OpenAI) for language improvement during the preparation of the manuscript.

%%%%%%%%%%%%%%%%%%%%%%%%%%%%%%%%%%%%%%
%%%%%%%%%%%%%%%%%%%%%%%%%%%%%%%%%%%%%%
\begin{widetext}
\begin{appendix}
\setcounter{equation}{0}
\setcounter{table}{0}
\makeatletter

\section*{Appendix}

\section{Cumulant equations of motion}
\label{app:cumulants}
Let us define a classical Hamiltonian functional as 
\begin{equation}
    \mathcal{H}=\lim_{L\rightarrow \infty} \frac{1}{L} \expval{H}\,,
\end{equation}
where the expectation value is taken in the subspace of $\mathrm{SU}(3)$ coherent states, which are Gaussian with respect to the bosonic bilinears and preserve the $\mathrm{U}(1)$ particle number symmetry of the model. We further make an additional assumption that the system is homogeneous, allowing us to factorize $L^{-1}\sum_{j=1}^L \expval{\hat{T}_{j,\alpha\beta}}\expval{\hat{T}_{j,\mu\nu}}\approx T_{\alpha\beta} T_{\mu\nu} $. Applying all the symmetry constraints, we can expand the classical Hamiltonian functional in the basis of the collective spins
\begin{align}
\label{eq:Hamiltonian_classical}
\mathcal{H}
&=
h\left(T_{11} - T_{33}\right)-\left(g_+T_{12} + g_-T_{23}\right)\left(g_+T_{21} + g_-T_{32}\right) -
\frac{1}{L}\left[g^2_+T_{11}\left(T_{22}+1\right) + g^2_-T_{22}\left(T_{33}+1\right) \right]\nonumber\\
&-
\frac{1}{L} g_+g_-\left(T_{23}T_{21} + T_{12}T_{32}\right) + \mathcal{O}\left(1/L^2\right). 
\end{align}
The corresponding semiclassical equations of motion are then dictated by the Poisson--Lie structure dictated by the Hamiltonian functional. The generators of the dynamics are the derivatives of the Hamiltonian functional with respect to the phase-space variable, i.e., the collective spins,
\begin{equation}
    \frac{\partial \mathcal{H}}{\partial T_{\alpha\beta}}=\tilde{M}_{\beta\alpha}\,,
\end{equation}
where up to the order $\mathcal{O}\left(1/L\right)$ we have
\begin{equation}
\label{eq:M_tilde}
\tilde{M}
=
-
\begin{pmatrix}
-h + \frac{g_{+}^2}{L}\left(T_{22} + 1\right) & g_{+}^2 T_{12} + g_{+} g_{-}\left(1 + \frac{1}{L}\right) T_{23} & 0 \\
 g_{+}^2 T_{21} + g_{+} g_{-}\left(1 + \frac{1}{L}\right) T_{32} & \frac{g_{+}^2}{L} T_{11} + \frac{g^2_{-}}{L}\left(T_{33} + 1\right) & g_{-}^2
T_{23} + g_{+}g_{-}\left(1 + \frac{1}{L}\right) T_{12} \\ 
0 & g_{-}^2
T_{32} + g_{+}g_{-}\left(1 + \frac{1}{L}\right) T_{21} & h + \frac{g_{-}^2}{L} T_{22} 
\end{pmatrix}.
\end{equation}
Using the fully antisymmetric nature of the $\mathrm{SU}(3)$ structure constant, the equations of motion can be written as 
\begin{equation}
\label{eq:dT_cumulants}
\im \frac{\dd}{\dd t} T(t)
=
[T(t),\tilde{M}(t)]\,,
\end{equation}
in a matrix form, or
\begin{equation}
\mathrm{i}\frac{\dd}{\dd t}T_{\alpha\beta}(t)=\{T_{\alpha\beta}(t),\mathcal{H}\}_{\mathrm{PB}}\,,
\end{equation}
in a Hamiltonian form, where the Poisson--Lie bracket is defined as
\begin{equation}
\{\ \cdot \ , \ \cdot \ \}_{\mathrm{PB}} =\sum_{\alpha,\beta,\gamma=1}^3 T_{\alpha\gamma}\left[ \frac{\overleftarrow{\partial}}{\partial T_{\alpha\beta}}\frac{\overrightarrow{\partial}}{\partial T_{\gamma\beta}} - \frac{\overleftarrow{\partial}}{\partial T_{\gamma\beta}}\frac{\overrightarrow{\partial}}{\partial T_{\alpha\beta}}\right].
\end{equation}

\section{$\mathrm{SU}(3)$ coherent states}
\label{app:coherent}
 $\mathrm{SU}(3)$ coherent states can be defined as \cite{Gnutzmann2000Quantum}
\begin{equation}
    \ket{\boldsymbol{\gamma}}_{p,q} = \frac{1}{\sqrt{A_1^pA_2^q}}\,\mathrm{e}^{\gamma_3 \hat{T}_{31}}\mathrm{e}^{\gamma_1 \hat{T}_{21}} \mathrm{e}^{\gamma_2 \hat{T}_{32}}\ket{\mu}_{p,q}\,,
\end{equation}
where $\gamma_i$ are free parameters, $\ket{\mu}_{p,q}$ is the highest weight state of the $D(p,q)$ representation of $\mathrm{SU}(3)$, and the normalization constants read $A_1 = 1 + |\gamma_1|^2 + |\gamma_3|^2$, $A_2 = 1 + |\gamma_2|^2 + |\gamma_3 -\gamma_1\gamma_2|^2$. The $\mathrm{SU}(3)$ spin expectation values for such states read
\begin{align}
T_{11} &= \frac{\mathcal{N}_a}{3} +\frac{p}{A_1}\left(2 - |\gamma_1|^2 - |\gamma_3|^2\right) + \frac{q}{3A_2}\left(1 + |\gamma_2|^2 - 2|\gamma_3 - \gamma_1 \gamma_2|^2\right),\nonumber\\
T_{22} &= \frac{\mathcal{N}_a}{3} +\frac{p}{A_1}\left(-1 +2 |\gamma_1|^2 - |\gamma_3|^2\right) + \frac{q}{3A_2}\left(1 - 2 |\gamma_2|^2 + |\gamma_3 - \gamma_1 \gamma_2|^2\right),\nonumber\\
T_{33} &= \frac{\mathcal{N}_a}{3} +\frac{p}{A_1}\left(-1 - |\gamma_1|^2 + 2|\gamma_3|^2\right) + \frac{q}{3A_2}\left(-2 + |\gamma_2|^2 + |\gamma_3 - \gamma_1 \gamma_2|^2\right),\nonumber\\
T_{12} &= \frac{p}{A_1} \gamma_1 - \frac{q}{A_2} \bar{\gamma}_2\left(\gamma_3 - \gamma_1 \gamma_2\right),\quad
T_{13} = \frac{p}{A_1}\gamma_3 + \frac{q}{A_2}\left(\gamma_3 - \gamma_1 \gamma_2\right),\quad
T_{23} = \frac{p}{A_1}\bar{\gamma}_1 \gamma_3 + \frac{q}{A_2}\gamma_2\,,
\end{align}
where $\mathcal{N}_a$ is the particle number, determined by the condition $\left(\mathcal{N}_a + 1\right)\left(\mathcal{N}_a + 2\right) = \left(p+1\right)\left(q + 1\right)\left(p+q+2\right)$.

As detailed in Ref.~\cite{Balducci:2025}, coherent states exhibit different dynamical responses depending on the representation they are defined in. In particular, states with $p=0$ or $q=0$ are strictly regular. In this work, we chose a representation $D(2,1)$, which implies $\mathcal{N}_a = 4$ particles per site. This is a representative state exhibiting chaotic dynamics in some parts of the parameter space. In the main text, we refer to it as a ``chaotic initial state''.
The exact parametrization chosen for this work is $\gamma_1=4/\sqrt{21}$, $\gamma_2=2/\sqrt{21}$, and $\gamma_3=\mathrm{i}/\sqrt{21}$. We emphasize that the precise choice of the parametrization does not affect the general qualitative features of the phase diagram depicted in Fig.~\ref{fig:phase_diagram}. A possible exception is extreme cases in which one or more of the parameters $\gamma_i$ are close to zero, effectively confining the dynamics to a smaller symmetry sector.

\section{2PI derivation}
\label{app:2PI}
The Schwinger--Keldysh action corresponding the Hamiltonian \eqref{eq:hamiltonian} takes the form 
\begin{align}
S[\mathfrak{b},\bar{\mathfrak{b}}]
&=
\int_{\mathcal{C}}\dd{t} \left[\frac{\im}{2}\left(\bar{\mathfrak{b}}_{i,\alpha} \partial_t \mathfrak{b}_{i,\alpha} - \mathfrak{b}_{i,\alpha} \partial_t \bar{\mathfrak{b}}_{i,\alpha}\right) + \left(\frac{g_{+}^2}{L}-h\right)\bar{\mathfrak{b}}_{i,1}\mathfrak{b}_{i,1} + \frac{g_{-}^2}{L}\bar{\mathfrak{b}}_{i,2} \mathfrak{b}_{i,2} + h\bar{\mathfrak{b}}_{i,3}\mathfrak{b}_{i,3}\right.\nonumber\\
&+
\left.\frac{1}{L} \left(g_{+}\bar{\mathfrak{b}}_{i,1}\mathfrak{b}_{i,2} + g_{-}\bar{\mathfrak{b}}_{i,2}\mathfrak{b}_{i,3}\right)\left(g_{+}\bar{\mathfrak{b}}_{j,2}\mathfrak{b}_{j,1} + g_{-}\bar{\mathfrak{b}}_{j,3}\mathfrak{b}_{j,2}\right)\right] .
\end{align}
Same as in the main text, the standard Einstein summation convention over repeated indices $i,j \in \lbrace 1,\ldots,L\rbrace$ and $\alpha \in \lbrace 1, 2, 3\rbrace$ is implied. Introducing vector notation, $\Phi_i = \left(\mathfrak{b}_{i,1}, \mathfrak{b}_{i,2}, \mathfrak{b}_{i,3}\right)$, this action can be compactly written as 
\begin{equation}
S[\Phi,\bar{\Phi}]
=
\int_{\mathcal{C}}\dd{t} \left[\frac{\im}{2}\left(\bar{\Phi}_i^{\alpha} \partial_t \Phi_i^{\alpha} - \Phi^{\alpha}_i \partial_t \bar{\Phi}^{\alpha}_i\right) + B^{\alpha} \bar{\Phi}_i^{\alpha} \Phi_i^{\alpha} + \frac{K^T_{\alpha\beta} K_{\gamma\delta}}{L} \bar{\Phi}_i^{\alpha}\Phi_i^{\beta} \bar{\Phi}_j^{\gamma}\Phi_j^{\delta}\right].
\end{equation}
where
\begin{equation}
B = \left(\frac{g_+^2}{L}-h, \frac{g_-^2}{L}, h\right) \quad \mathrm{and}\quad K 
=
\begin{pmatrix}
0     & 0     & 0\\
g_{+} & 0     & 0\\
0     & g_{-} & 0
\end{pmatrix}\,.
\end{equation}
The interaction term can be decoupled via the Hubbard--Stratonovich transformation to wit:
\begin{equation}
\label{eq:S}
S[\Phi,\bar{\Phi},\psi,\bar{\psi}]
=
\int_{\mathcal{C}}\dd{t} \left[\frac{\im}{2}\left(\bar{\Phi}_i^{\alpha} \partial_t \Phi_i^{\alpha} - \Phi^{\alpha}_i \partial_t \bar{\Phi}^{\alpha}_i\right) + B_{\alpha} \bar{\Phi}_i^{\alpha} \Phi_i^{\alpha} - \bar{\psi} \psi
+ \frac{1}{\sqrt{L}} \left(\psi K^T_{\alpha\beta} + \bar{\psi} K_{\alpha\beta}\right) \bar{\Phi}_i^{\alpha}\Phi_i^{\beta}\right].
\end{equation}
The two-particle irreducible (2PI) effective action can be written in the standard form
\begin{equation}
\label{eq:Gamma_2PI}
\Gamma_{\mathrm{2PI}} = S + \frac{\im}{2} \Tr_{\mathcal{C}}\ln G^{-1} + \frac{\im}{2}
\Tr_{\mathcal{C}}\ln D^{-1} + \frac{\im}{2} \Tr_{\mathcal{C}} G_0^{-1}\,G + \frac{\im}{2}
\Tr_{\mathcal{C}} D_0^{-1}\,D + \Gamma_2.
\end{equation}
where $\Gamma_2 = -\im\ln\langle\exp(\im S_{\mathrm{int}})\rangle_{\mathrm{2PI}}$ is the sum of all two-particle irreducible, with respect to the full propagators $G$ and $D$, connected vacuum diagrams and
\begin{equation}
\label{eq:bare_prop}
\left(G_0^{-1}\right)_{ab,ij}^{\alpha\beta}(t,t')
=
\delta_{ij}\delta_{\mathcal{C}}(t-t')\left[\sigma^z_{ab} \delta^{\alpha\beta}\partial_{t'} + \im
\begin{pmatrix}
m^{\alpha\beta} & 0 \\
0 & m^{\beta\alpha}
\end{pmatrix}_{ab}
\right],\quad
D_{0,ab}^{-1}(t,t') 
=
\im\delta_{ab}\delta_{\mathcal{C}}(t-t')
\end{equation}
are the the classical inverse propagators for the bosonic $\Phi$ and the Hubbard–Stratonovich $\psi$ fields, respectively. Here,
\begin{equation}
m_{\alpha\beta} = -B_{\alpha}\delta_{\alpha\beta} - \frac{\psi K_{\alpha\beta}^T + \bar{\psi}K_{\alpha\beta}}{\sqrt{L}}\,,
\end{equation}
and we have adopted the standard convention
\begin{equation}
D_{ab}(t,t') = \langle\mathcal{T}_C \hat{\psi}_a(t)\,\hat{\psi}_{\bar{b}}(t')\rangle_c 
=
\begin{pmatrix}
D_{\psi\bar{\psi}}(t,t') & D_{\psi\psi}(t,t') \\
D_{\bar{\psi}\bar{\psi}}(t,t') & D_{\bar{\psi}\psi}(t,t')  
\end{pmatrix}\,,
\end{equation}
where the indices $a,b \in \lbrace 1, 2\rbrace$, with $\bar{a} \equiv 3 - a$, and $\hat{\psi}_1(t) \equiv \hat{\psi}(t)$, $\hat{\psi}_2(t) \equiv \hat{\psi}^{\dagger}(t)$, and likewise for $G$. 

One advantage of performing the Hubbard--Stratonovich transformation is that it allows to efficiently resum an infinite number of diagrams at a finite-loop order. For example, introducing the diagrammatic notation
\begin{equation*}
\frac{1}{\sqrt{L}} K^{(T)} \int_{\mathcal{C}}\dd{t} 
\ = \
\begin{tikzpicture}[anchor=base, baseline=-0.1cm]
\filldraw (-4ex,0) circle (0.3ex);
\draw[thick] (-6ex,3ex) -- (-4ex,0);
\draw[thick] (-6ex,-3ex) -- (-4ex,0);
\draw[style=dotted,very thick] (-4ex,0) to (2ex,0);
\end{tikzpicture}
\hspace{5ex}
D(t,t')
\ = \
\begin{tikzpicture}[anchor=base, baseline=-0.1cm]
\draw[style=dotted,very thick] (-4ex,0) to (4ex,0);
\end{tikzpicture}
\hspace{5ex}
G(t,t')
\ = \
\begin{tikzpicture}[anchor=base, baseline=-0.1cm]
\draw[thick] (-4ex,0) to (4ex,0);
\end{tikzpicture}
\end{equation*}
the infinite sum of the $O(L^0)$ bubble-chain diagrams, Fig.~\ref{fig:diagrams}(c), discussed in the main text, is equivalent to the following two-loop diagram:
\begin{align}
\label{eq:NLO}
\Gamma_2^{\mathrm{NLO}} 
\; = \;
\begin{tikzpicture}[anchor=base, baseline=-0.1cm]
\filldraw (-4ex,0) circle (0.3ex);
\filldraw (4ex,0) circle (0.3ex);
\draw[thick] (0,0) circle (4ex);
\draw[dotted,very thick] (-4ex,0) to (4ex,0);
\end{tikzpicture}
\; = \;
\frac{\im}{2L}\int_{\mathcal{C}}\dd{t}\dd{t'}&\left[K^T_{\alpha\beta} K^T_{\delta\gamma} D_{12}(t,t') + K^T_{\alpha\beta} K_{\delta\gamma} D_{11}(t,t') + K_{\alpha\beta} K^T_{\delta\gamma} D_{22}(t,t') + K_{\alpha\beta} K_{\delta\gamma} D_{21}(t,t')\right]\nonumber\\ 
&\times
\left[G_{22,ij}^{\alpha\delta}(t,t')\,G_{11,ij}^{\beta\gamma}(t,t') + G_{21,ij}^{\alpha\gamma}(t,t')\,G_{12,ij}^{\beta\delta}(t,t')\right].
\end{align}
The quantum equations of motion for the one- and two-point functions are obtained by extremizing the 2PI effective action \eqref{eq:Gamma_2PI}:
\begin{equation}
\frac{\delta \Gamma_{\mathrm{2PI}}}{\delta \Phi} = 0\,,\quad 
\frac{\delta \Gamma_{\mathrm{2PI}}}{\delta \psi} = 0\,,\quad 
\frac{\delta \Gamma_{\mathrm{2PI}}}{\delta G} = 0\,,\quad
\frac{\delta \Gamma_{\mathrm{2PI}}}{\delta D} = 0\,.
\end{equation}
In this work, we restrict ourselves to $\mathrm{U}(1)$-symmetric states such that $\langle\hat{\Phi}\rangle = 0$. The Hubbard--Stratonovich one-point function, on the other hand, is $\mathrm{U}(1)$-invariant, with its equation reading
\begin{equation}
\langle\hat{\psi}(t)\rangle = \frac{1}{\sqrt{L}} K_{\alpha\beta} G_{22,ii}^{\alpha\beta}(t,t)\,,\quad 
\langle\hat{\psi}^{\dagger}(t)\rangle = \frac{1}{\sqrt{L}} K^T_{\alpha\beta} G_{22,ii}^{\alpha\beta}(t,t)\,.
\end{equation}
Similarly, variation of \eqref{eq:Gamma_2PI} with respect to $G$ and $D$ yields
\begin{align}
\label{eq:DysonEq}
&\int_{\mathcal{C}}\dd{t''} \left[\left(G_0^{-1}\right)_{ac,ik}^{\alpha\gamma}(t,t'') - \Sigma_{ac,ik}^{\alpha\gamma}(t,t'')\right] G_{kj,cb}^{\gamma\beta}(t'',t') = \delta_{ab}\delta_{ij}\delta^{\alpha\beta}\delta_{\mathcal{C}}(t - t')\,,\\
&\int_{\mathcal{C}}\dd{t''} \left[D_{0,ac}^{-1}(t,t'') - \Omega_{ac}(t,t'')\right] D_{cb}(t'',t') = \delta_{ab}\delta_{\mathcal{C}}(t - t')\,,
\end{align}
with the proper self-energies $\Sigma$ and $\Omega$ defined as
\begin{equation}
\Sigma_{ab,ij}^{\alpha\beta}(t,t') = 2\im\frac{\delta \Gamma_2[G,D]}{\delta G_{ba,ji}^{\beta\alpha}(t',t)}\,,\quad 
\Omega_{ab}(t,t') = 2\im\frac{\delta \Gamma_2[G,D]}{\delta D_{ba}(t',t)}\,.
\end{equation}
The action \eqref{eq:S} is invariant under on-site $\mathrm{U}(1)$ transformations $\Phi^{\alpha}_i(t) \to \exp\left(\im\varepsilon_i\right) \Phi^{\alpha}_i(t)$, $\bar{\Phi}^{\alpha}_i(t) \to \exp\left(-\im\varepsilon_i\right) \bar{\Phi}^{\alpha}_i(t)$. Assuming this symmetry is not broken, this implies that $\Sigma_{ab,ij}^{\alpha\beta} = \delta_{ij}\delta_{ab}\Sigma_{a,i}^{\alpha\beta}$ and $G_{ab,ij}^{\alpha\beta} = \delta_{ij}\delta_{ab}G_{a,i}^{\alpha\beta}$. In this work, we focus on the spatially homogeneous case, for which $\Sigma_{a,i}^{\alpha\beta} = \Sigma_a^{\alpha\beta}$ and $G_{a,i}^{\alpha\beta} = G_a^{\alpha\beta}$. In addition, we will adopt the notation $G = G^{b \bar{b}}$, $\tilde{G} = G^{\bar{b}b}$, $\Sigma = \Sigma^{b \bar{b}}, \tilde{\Sigma} = \Sigma^{\bar{b}b}$ for the only nonvanishing components. With these conventions in mind, the $O(1/L)$ self-energies can be written as
\begin{align}
\label{eq:self_energies}
\Sigma_{\alpha\beta}(t,t')
&=
-\frac{1}{L}\left[K^T_{\alpha\gamma} D_{12}(t,t') K^T_{\delta\beta} + K_{\alpha\gamma} D_{22}(t,t') K^T_{\delta\beta} + 
K^T_{\alpha\gamma} D_{11}(t,t') K_{\delta\beta} + K_{\alpha\gamma} D_{21}(t,t') K_{\delta\beta} \right]G_{\gamma\delta}(t,t')\,,\nonumber\\
\tilde{\Sigma}_{\alpha\beta}(t,t')
&=
-\frac{1}{L}\left[K_{\alpha\gamma} D_{12}(t,t') K_{\delta\beta} + K^T_{\alpha\gamma} D_{22}(t,t') K_{\delta\beta} + 
K_{\alpha\gamma} D_{11}(t,t') K^T_{\delta\beta} + K^T_{\alpha\gamma} D_{21}(t,t') K^T_{\delta\beta} \right] \tilde{G}_{\gamma\delta}(t,t')
=
\Sigma_{\beta\alpha}(t',t)\,,\\
\Omega(t,t')
&=
-\frac{1}{L}
\begin{pmatrix}
K_{\gamma\delta} G_{\delta\alpha}(t,t') K^T_{\alpha\beta}\tilde{G}^T_{\beta\gamma}(t,t')  &  K_{\gamma\delta} G_{\delta\alpha}(t,t') K_{\alpha\beta}\tilde{G}^T_{\beta\gamma}(t,t') \\
K^T_{\gamma\delta} G_{\delta\alpha}(t,t') K^T_{\alpha\beta}\tilde{G}^T_{\beta\gamma}(t,t')  & K^T_{\gamma\delta} G_{\delta\alpha}(t,t') K_{\alpha\beta}\tilde{G}^T_{\beta\gamma}(t,t') 
\end{pmatrix}\,.
\end{align}

The nonequilibrium Dyson equations \eqref{eq:DysonEq} for correlation functions on the Schwinger--Keldysh contour can be conveniently mapped to equations on the regular time axis by decomposing the self-energies and propagators as
\begin{align}
D(t,t') &= D_0(t,t') + D^F(t,t') - \frac{\im}{2}\mathrm{sgn}_C(t-t')\,D^{\rho}(t,t')\,,\quad
\Omega(t,t') = \Omega^{(0)}(t,t')\,\delta_{\mathcal{C}}(t-t') + \Omega^F(t,t') - \frac{\im}{2}\mathrm{sgn}_C(t-t')\,\Omega^{\rho}(t,t')\,,\nonumber\\
G(t,t') &= F(t,t') - \frac{\im}{2}\mathrm{sgn}_C(t-t')\,\rho(t,t')\,,\quad
\Sigma(t,t') = \Sigma^{(0)}(t)\,\delta_{\mathcal{C}}(t-t') + \Sigma^F(t,t') - \frac{\im}{2}\mathrm{sgn}_C(t-t')\,\Sigma^{\rho}(t,t')\,.
\end{align}
Here, we explicitly separated the singular part of the Hubbard--Stratonovich propagator given by its bare component $D_0(t,t') \propto \delta_{\mathcal{C}}(t-t')$, cf. Eq.~\eqref{eq:bare_prop}. The resulting equations are given by
\begin{align}
\label{eq:KB_G}
\im\partial_t F_{\alpha\beta}(t,t') 
&=
M_{\alpha\gamma}(t) F_{\gamma\beta}(t,t')
+
\int_{t_0}^t\dd{t''}\Sigma^{\rho}_{\alpha\gamma}(t,t'') F_{\gamma\beta}(t'',t') 
-
\int_{t_0}^{t'}\dd{t''}\Sigma^F_{\alpha\gamma}(t,t'') \rho_{\gamma\beta}(t'',t')\,,\nonumber\\
\im\partial_t \rho_{\alpha\beta}(t,t') 
&=
M_{\alpha\gamma}(t) \rho_{\gamma\beta}(t,t')
+
\int_{t'}^t\dd{t''}\Sigma^{\rho}_{\alpha\gamma}(t,t'') \rho_{\gamma\beta}(t'',t')\,,\\
\label{eq:KB_Gtilde}
-\im\partial_t \tilde{F}_{\alpha\beta}(t,t') 
&=
\tilde{M}_{\alpha\gamma}(t) \tilde{F}_{\gamma\beta}(t,t')
+
\int_{t_0}^t\dd{t''}\tilde{\Sigma}^{\rho}_{\alpha\gamma}(t,t'') \tilde{F}_{\gamma\beta}(t'',t') 
-
\int_{t_0}^{t'}\dd{t''}\tilde{\Sigma}^F_{\alpha\gamma}(t,t'') \tilde{\rho}_{\gamma\beta}(t'',t')\,,\nonumber\\
-\im\partial_t \tilde{\rho}_{\alpha\beta}(t,t') 
&=
\tilde{M}_{\alpha\gamma}(t) \tilde{\rho}_{\gamma\beta}(t,t')
+
\int_{t'}^t\dd{t''}\tilde{\Sigma}^{\rho}_{\alpha\gamma}(t,t'') \tilde{\rho}_{\gamma\beta}(t'',t')\,,
\end{align}
and
\begin{align}
\label{eq:HS_bethesalpeter}
D^F_{ab}(t,t')
&=
-\Omega^F_{ab}(t,t') - \int_{t_0}^t\dd{t''}\Omega^{\rho}_{ac}(t,t'') D^F_{cb}(t'',t') + \int_{t_0}^{t'}\dd{t''}\Omega^F_{ac}(t,t'') D^{\rho}_{cb}(t'',t'),\nonumber\\
D^{\rho}_{ab}(t,t')
&=
-\Omega^{\rho}_{ab}(t,t') - \int_{t'}^t\dd{t''}\Omega^{\rho}_{ac}(t,t'')D^{\rho}_{cb}(t'',t')\,,
\end{align}
with
\begin{equation}
\label{eq:M_2PI}
M(t)
=
m(t)
+
\im\Sigma^{(0)}(t)
=
-
\begin{pmatrix}
-h + \frac{g_{+}^2}{L}\left(F_{22} + 1\right) & g_{+}^2 F_{12} + g_{+} g_{-}\left(1 + \frac{1}{L}\right) F_{23} & 0 \\
 g_{+}^2 F_{21} + g_{+} g_{-}\left(1 + \frac{1}{L}\right) F_{32} & \frac{g_{+}^2}{L} F_{11} + \frac{g^2_{-}}{L}\left(F_{33} + 1\right) & g_{-}^2
F_{23} + g_{+}g_{-}\left(1 + \frac{1}{L}\right) F_{12} \\ 
0 & g_{-}^2
F_{32} + g_{+}g_{-}\left(1 + \frac{1}{L}\right) F_{21} & h + \frac{g_{-}^2}{L} F_{22} 
\end{pmatrix}
=
\tilde{M}(t)^T,    
\end{equation}
Here, all the temporal arguments on the right-hand side assumed to be $(t,t)$. Note that $\tilde{M}$ is the same object as derived in the semiclassical computation in equation \eqref{eq:M_tilde}. Finally, decomposition of the self-energies \eqref{eq:self_energies} reads
\begin{align}
\label{eq:self_energies1}
\Sigma^F_{\alpha\beta}
&=
-\frac{1}{L}\left[K^T_{\alpha\gamma} D^F_{12} K^T_{\delta\beta} + K_{\alpha\gamma} D^F_{22} K^T_{\delta\beta} + 
K^T_{\alpha\gamma} D^F_{11} K_{\delta\beta} + K_{\alpha\gamma} D^F_{21} K_{\delta\beta} \right] F_{\gamma\delta}\nonumber\\
&+
\frac{1}{4 L}\left[K^T_{\alpha\gamma} D^{\rho}_{12} K^T_{\delta\beta} + K_{\alpha\gamma} D^{\rho}_{22} K^T_{\delta\beta} + 
K^T_{\alpha\gamma} D^{\rho}_{11} K_{\delta\beta} + K_{\alpha\gamma} D^{\rho}_{21} K_{\delta\beta} \right] \rho_{\gamma\delta}\,,\nonumber\\
\Sigma^{\rho}_{\alpha\beta}
&=
-\frac{1}{L}\left[K^T_{\alpha\gamma} D^F_{12} K^T_{\delta\beta} + K_{\alpha\gamma} D^F_{22} K^T_{\delta\beta} + 
K^T_{\alpha\gamma} D^F_{11} K_{\delta\beta} + K_{\alpha\gamma} D^F_{21} K_{\delta\beta} \right] \rho_{\gamma\delta}\nonumber\\
&-
\frac{1}{L}\left[K^T_{\alpha\gamma} D^{\rho}_{12} K^T_{\delta\beta} + K_{\alpha\gamma} D^{\rho}_{22} K^T_{\delta\beta} + 
K^T_{\alpha\gamma} D^{\rho}_{11} K_{\delta\beta} + K_{\alpha\gamma} D^{\rho}_{21} K_{\delta\beta} \right] F_{\gamma\delta}\,,\nonumber\\
\tilde{\Sigma}^F_{\alpha\beta}
&=
-\frac{1}{L}\left[K_{\alpha\gamma} D^F_{12} K_{\delta\beta} + K^T_{\alpha\gamma} D^F_{22} K_{\delta\beta} + 
K_{\alpha\gamma} D^F_{11} K^T_{\delta\beta} + K^T_{\alpha\gamma} D^F_{21} K^T_{\delta\beta} \right] \tilde{F}_{\gamma\delta}\nonumber\\
&+
\frac{1}{4L}\left[K_{\alpha\gamma} D^{\rho}_{12} K_{\delta\beta} + K^T_{\alpha\gamma} D^{\rho}_{22} K_{\delta\beta} + 
K_{\alpha\gamma} D^{\rho}_{11} K^T_{\delta\beta} + K^T_{\alpha\gamma} D^{\rho}_{21} K^T_{\delta\beta} \right] \tilde{\rho}_{\gamma\delta}\,,\nonumber\\
\tilde{\Sigma}^{\rho}_{\alpha\beta}
&=
-\frac{1}{L}\left[K_{\alpha\gamma} D^F_{12} K_{\delta\beta} + K^T_{\alpha\gamma} D^F_{22} K_{\delta\beta} + 
K_{\alpha\gamma} D^F_{11} K^T_{\delta\beta} + K^T_{\alpha\gamma} D^F_{21} K^T_{\delta\beta} \right] \tilde{\rho}_{\gamma\delta}\nonumber\\
&-
\frac{1}{L}\left[K_{\alpha\gamma} D^{\rho}_{12} K_{\delta\beta} + K^T_{\alpha\gamma} D^{\rho}_{22} K_{\delta\beta} + 
K_{\alpha\gamma} D^{\rho}_{11} K^T_{\delta\beta} + K^T_{\alpha\gamma} D^{\rho}_{21} K^T_{\delta\beta} \right] \tilde{F}_{\gamma\delta}\,,\\
\label{eq:self_energies2}
\Omega^F
&=
-\frac{1}{L}
\begin{pmatrix}
K_{\gamma\delta} F_{\delta\alpha} K^T_{\alpha\beta}\tilde{F}^T_{\beta\gamma}  &  K_{\gamma\delta} F_{\delta\alpha} K_{\alpha\beta}\tilde{F}^T_{\beta\gamma} \\
K^T_{\gamma\delta} F_{\delta\alpha} K^T_{\alpha\beta}\tilde{F}^T_{\beta\gamma}  & K^T_{\gamma\delta} F_{\delta\alpha} K_{\alpha\beta}\tilde{F}^T_{\beta\gamma} 
\end{pmatrix}
+
\frac{1}{4L}
\begin{pmatrix}
K_{\gamma\delta} \rho_{\delta\alpha} K^T_{\alpha\beta}\tilde{\rho}^T_{\beta\gamma}  &  K_{\gamma\delta} \rho_{\delta\alpha} K_{\alpha\beta}\tilde{\rho}^T_{\beta\gamma} \\
K^T_{\gamma\delta} \rho_{\delta\alpha} K^T_{\alpha\beta}\tilde{\rho}^T_{\beta\gamma}  & K^T_{\gamma\delta} \rho_{\delta\alpha} K_{\alpha\beta}\tilde{\rho}^T_{\beta\gamma} 
\end{pmatrix}\,\nonumber\\
\Omega^{\rho}
&=
-\frac{1}{L}
\begin{pmatrix}
K_{\gamma\delta} F_{\delta\alpha} K^T_{\alpha\beta}\tilde{\rho}^T_{\beta\gamma}  &  K_{\gamma\delta} F_{\delta\alpha} K_{\alpha\beta}\tilde{\rho}^T_{\beta\gamma} \\
K^T_{\gamma\delta} F_{\delta\alpha} K^T_{\alpha\beta}\tilde{\rho}^T_{\beta\gamma}  & K^T_{\gamma\delta} F_{\delta\alpha} K_{\alpha\beta}\tilde{\rho}^T_{\beta\gamma} 
\end{pmatrix}
-
\frac{1}{L}
\begin{pmatrix}
K_{\gamma\delta} \rho_{\delta\alpha} K^T_{\alpha\beta}\tilde{F}^T_{\beta\gamma}  &  K_{\gamma\delta} \rho_{\delta\alpha} K_{\alpha\beta}\tilde{F}^T_{\beta\gamma} \\
K^T_{\gamma\delta} \rho_{\delta\alpha} K^T_{\alpha\beta}\tilde{F}^T_{\beta\gamma}  & K^T_{\gamma\delta} \rho_{\delta\alpha} K_{\alpha\beta}\tilde{F}^T_{\beta\gamma} 
\end{pmatrix}\,,
\end{align}
where, for brevity, the temporal arguments have been suppressed. 

Equations \eqref{eq:KB_G}, \eqref{eq:KB_Gtilde}, and \eqref{eq:HS_bethesalpeter} together with the self-energies \eqref{eq:self_energies1} and \eqref{eq:self_energies2} constitute a closed system of equations, which we solve numerically. The dynamics of the $\mathrm{SU}(3)$ spin variables can be readily extracted from the equal-time statistical propagators by noticing $T_{\alpha\beta}(t) = \tilde{F}_{\alpha\beta}(t,t) - \delta_{\alpha\beta}/2$, so that
\begin{align}
\im\frac{\dd}{\dd t} T_{\alpha\beta}(t) 
=
\left.\im\left(\partial_t + \partial_t'\right) \tilde{F}_{\alpha\beta}(t,t')\right\rvert_{t=t'}
=
T_{\alpha\gamma}(t)\tilde{M}_{\gamma\beta}(t) - \tilde{M}_{\alpha\gamma}(t)T_{\gamma\beta}(t) 
&-
\int_{t_0}^t \dd{t''}\left[\tilde{\Sigma}^{\rho}_{\alpha\gamma}(t,t'') \tilde{F}_{\gamma\beta}(t'',t) 
-
\tilde{\Sigma}^F_{\alpha\gamma}(t,t'') \tilde{\rho}_{\gamma\beta}(t'',t)\right] \nonumber\\
&+
\int_{t_0}^t \dd{t''}\left[\tilde{F}_{\alpha\gamma}(t,t'') \tilde{\Sigma}^{\rho}_{\gamma\beta}(t'',t)
-
\tilde{\rho}_{\alpha\gamma}(t,t'')\tilde{\Sigma}^F_{\gamma\beta}(t'',t) \right],
\end{align}
or in compact form,
\begin{equation}
\im\frac{\dd}{\dd t} T(t)
=
[T(t),\tilde{M}(t)] + \int_{t_0}^t \dd{t''}\left[\tilde{F}(t,t'') \tilde{\Sigma}^{\rho}(t'',t)  - \tilde{\Sigma}^{\rho}(t,t'')\tilde{F}(t'',t)
-
\tilde{\rho}(t,t'')\tilde{\Sigma}^F(t'',t) + \tilde{\Sigma}^F(t,t'')\tilde{\rho}(t'',t)\right].
\end{equation}
Notice that the local-in-time part of the equation of motion is identical to Eq.~\eqref{eq:dT_cumulants} derived in the cumulant expansion, up to a subleading correction between the diagonal elements $T_{\alpha\alpha}$ and $F_{\alpha\alpha}$ in the matrix $\tilde{M}$. This discrepancy can be traced to the well-known operator-ordering subtlety that arises in the coherent-state path-integral formalism \cite{Polkovnikov:2009ys,Wilson:2010qbn}. In practice, the resulting $O(1/L)$ correction affects only a few matrix elements and has no significant impact on the dynamics. For definiteness, we therefore used the same expression \eqref{eq:M_2PI} for the matrix $\tilde{M}$ in both the 2PI and cumulant equations. 
\section{Dynamics of Coherences}
\label{app:coherences}
In Fig.~\ref{fig:coherences}, we plot the time evolution of the coherences for the collective $\mathrm{SU}(3)$ spins initially considered in Fig.~\ref{fig:observables}. Here, one can again compare the dynamical responses as computed from the 2PI and the cumulant equations of motion. 2PI suggests that the $T_{13}$ coherence relaxes, and the remainder of the coherences show persistent oscillations. On the other hand, the cumulant expansion methods show persistent oscillations for all off-diagonal correlation functions.

\newpage

\begin{center}
\begin{figure}[h]
    \includegraphics[width=\textwidth]{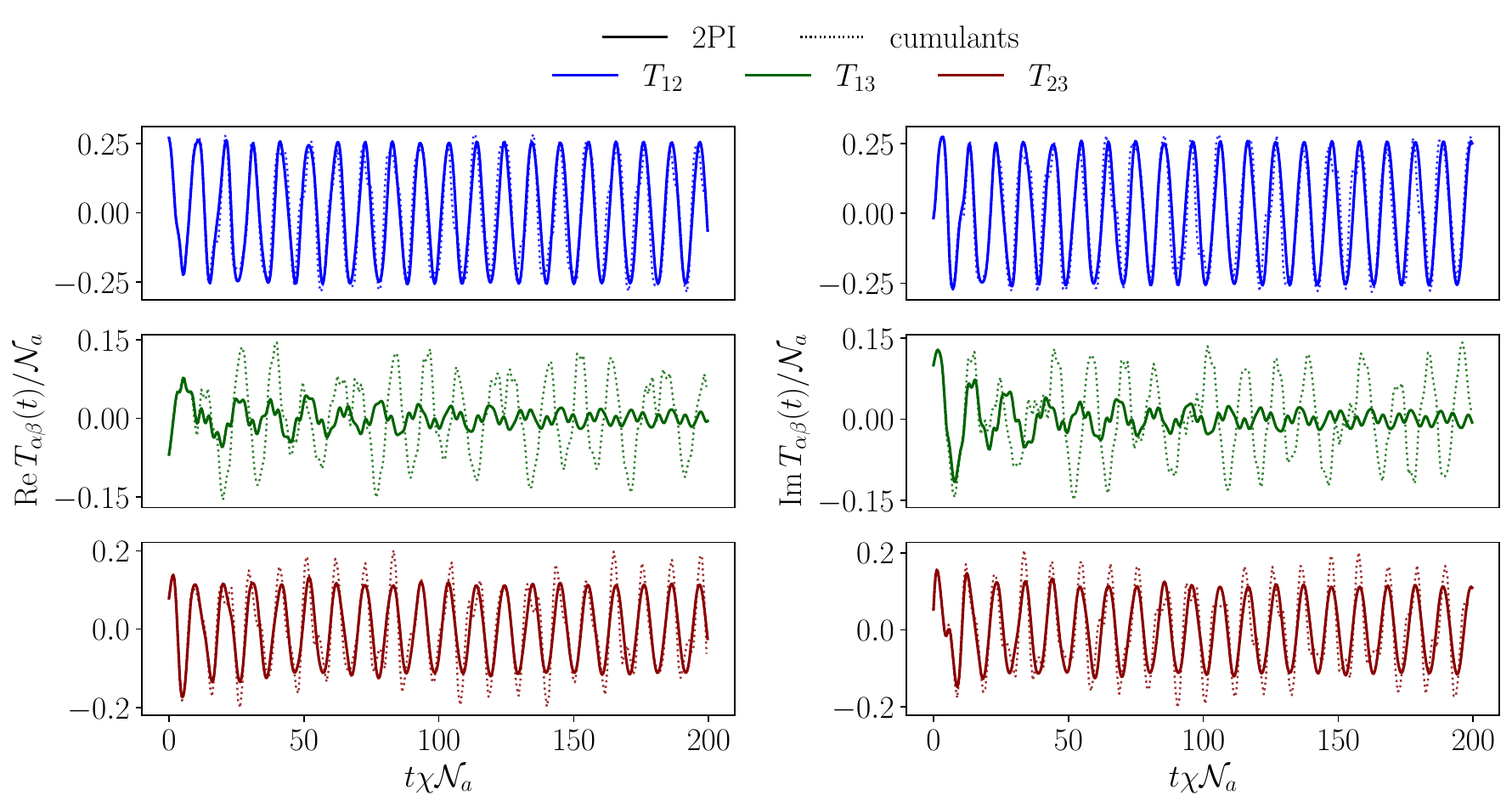}
    \caption{Time dependence of the off-diagonal two-point bosonic correlations for chaotic initial conditions, Hamiltonian parameters $h=1, g_+=2, g_-=1$, and the system size $L=15$.  We plot the time-evolution of the real part of the coherences (left panel) and the imaginary part of the coherences (right panel) as computed within the 2PI and the cumulant expansion approach. The 2PI calculations indicate relaxation of one of the observables, whereas the cumulant expansion shows persistent oscillations for every quantity considered.}
    \label{fig:coherences}
\end{figure}
\end{center}
\end{appendix}
\end{widetext}

\bibliographystyle{apsrev4-2}
\bibliography{bibliography}

\end{document}